# Cavity design for high-frequency axion dark matter detectors

I. Stern,[1] A.A. Chisholm,[1] J. Hoskins,[1] P. Sikivie,[1] N.S. Sullivan,[1] D.B. Tanner,[1] G. Carosi,[2] and K. van Bibber[3]

[1]*Department of Physics, University of Florida, Gainesville, Florida, 32611, USA*
[2]*Lawrence Livermore National Laboratory, Livermore, California, 94550, USA*
[3]*Department of Nuclear Engineering, University of California, Berkeley, California 94720, USA*

In an effort to extend the usefulness of microwave cavity detectors to higher axion masses, above ~8 μeV (~2 GHz), a numerical trade study of cavities was conducted to investigate the merit of using variable periodic post arrays and regulating vane designs for higher-frequency searches. The results show that both designs could be used to develop resonant cavities for high-mass axion searches. Multiple configurations of both methods obtained the scanning sensitivity equivalent to approximately 4 coherently coupled cavities with a single tuning rod.

## I. BACKGROUND

The detection of the axion is of significant interest to particle physics and to cosmology. The axion is a hypothetical pseudoscalar particle which was originally postulated to explain, within the framework of the standard model of particle interactions, why P (parity) and CP (charge conjugation times parity) are conserved by the strong interactions.[1-3] If the axion mass were in the range of $10^{-6}$-$10^{-3}$ eV, the axion is a natural cold dark matter (CDM) candidate.[4] With recent diminutions of the available parameter space for the weakly interacting massive particle (WIMP),[5] the axion is now by far the most promising candidate for the constitution of cold dark matter and this has generated a proliferation of searches around the world. String theory further suggests the simultaneous presence of many ultra-light axion-like particles, possibly populating each decade of mass down to the Hubble scale of $10^{-33}$ eV.[6]

Axion detection has proven to be extremely challenging. CDM axions have a lifetime vastly greater than the age of the universe,[5] have exceptionally weak interactions with matter and radiation, and were originally thought to be "invisible" to all detection techniques. However, in 1983 Sikivie[7] proposed a method by which these axions plausibly could be detected. He showed the decay rate of axions to photons can be greatly increased within a strong magnetic field, through the inverse Primakoff[8] effect. The Lagrangian for the axion-photon interaction is given by

$$\mathcal{L}_{a\gamma\gamma} = g_{a\gamma\gamma}\, a\, \mathbf{E} \cdot \mathbf{B}, \tag{1}$$

where $g_{a\gamma\gamma}$ is the axion-photon coupling constant, proportional to the mass of the axion, $a$ is the axion field, and $\mathbf{E}$ and $\mathbf{B}$ are the electric and magnetic fields, respectively. The coupling allows the axion to decay to two photons, as shown in Fig. 1(a). In a static external magnetic field, an axion may convert to a photon whose energy equals the total energy of the axion, as shown in Fig. 1(b). The $\mathbf{B}$ in Eq. (1) is effectively changed to the static magnetic field, $\mathbf{B_0}$. Thus, as the external magnetic field strength is increased, so is the conversion rate of the axion. This process is effective for relativistic or non-relativistic axions.[9]

The cold dark matter in the Milky Way halo has a velocity much less than the speed of light ($v \ll c$); if virialized,

$0 < v < c/1000$, making the spread of energy ~$10^{-6}\, mc^2$. The axion haloscope uses the inverse Primakoff effect to search for CDM axions in our galactic halo.

In the Sikivie[7] haloscope detector, a microwave cavity permeated by a strong magnetic field leads to resonant conversion of axions to photons. From Eq. (1), it can be shown that the coupling strength of the axion to a resonant mode is proportional to $\int d^3x\, \mathbf{B_0} \cdot \mathbf{E_{mnp}(x)}$, where $\mathbf{E_{mnp}}$ is the electric field of the mode. The indices $m$, $n$, and $p$ identify the various modes. For cylindrical cavities, they correspond to the polar coordinates $\phi$, $\rho$, and $z$, respectively. The integral is over the volume of the cavity.

The power produced in the cavity for a particular mode is given by

$$P_{mnp} \approx g_{a\gamma\gamma}{}^2 \frac{\rho_a}{m_a} B_0{}^2 V C_{mnp} Q_L, \tag{2}$$

where $m_a$ is the mass of the axion, $\rho_a$ is the local mass density of the axion field, $V$ is the volume of the cavity, and $Q$ is the quality factor of the cavity (assumed to be less than the kinetic energy spread of the axion at the Earth).[10] $C_{mnp}$ is the normalized form factor describing the coupling of the axion to a specific mode. It is given by

$$C_{mnp} \equiv \frac{\left| \int d^3x\, \mathbf{B_0} \cdot \mathbf{E_{mnp}(x)} \right|^2}{B_0{}^2 V \int d^3x\, \varepsilon(\mathbf{x}) |\mathbf{E_{mnp}(x)}|^2}, \tag{3}$$

where $\varepsilon(\mathbf{x})$ is the permittivity within the cavity normalized to vacuum.

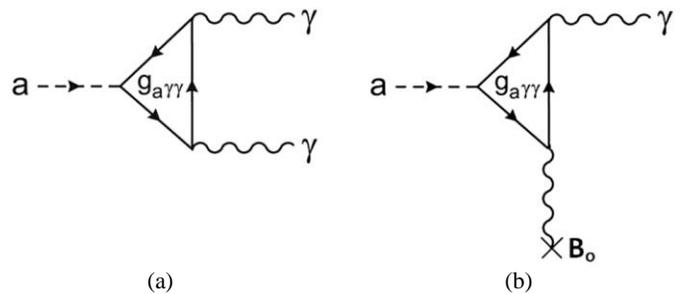

FIG. 1. Feynman diagrams for axion decay into photons. (a) The conversion in vacuum. (b) The inverse Primakoff effect in a static magnetic field ($\mathbf{B_0}$).

The mass of CDM axions is constrained by cosmology ($m_a \gtrsim$ μeV)[11-13] and astronomical observations ($m_a \lesssim 30$ meV).[14] The axion-photon coupling is model-dependent. In the KSVZ (Kim-Shifman-Vainshtein-Zakharov)[15,16] model, the coupling is

$$g_{a\gamma\gamma}^{KSVZ} = 0.38 \frac{m_a}{\text{GeV}^2} . \qquad (4)$$

In the DFSZ (Dine-Fischler-Srednicki-Zhitnitskii)[17,18] model,

$$g_{a\gamma\gamma}^{DFSZ} = 0.14 \frac{m_a}{\text{GeV}^2} . \qquad (5)$$

In all models, including the two above, the coupling is proportional to the axion mass but the numerical constant that appears generally differs by similar factors among the models. In all models where the Standard Model gauge group is grand-unified into a simple gauge group, the coupling is the same as in the DFSZ model.

## II. AXION DARK MATTER EXPERIMENT (ADMX)

The largest and most sensitive microwave cavity axion search to date is the Axion Dark Matter eXperiment (ADMX). ADMX searches for CDM axions using a 7.6 Tesla superconducting solenoid and a ~0.15 m³ cylindrical microwave resonator.[19] The magnetic field of the solenoid is aligned with the cylinder axis of the cavity (defined as the z-axis).

From Eq. (3), it can be seen that a signal will be produced only when the resonant mode's electric field integrates to a total nonzero z-component. Thus, all transverse electric modes (TE and TEM) will yield no detectable signal; only TM$_{0n0}$ modes will couple to axions. Further, the TM$_{010}$ mode has the largest form factor, with higher mode (TM$_{020}$, TM$_{030}$, TM$_{040}$, etc.) form factors decreasing as $\sim 1/f^2$ for a simple cylindrical cavity, with $f$ being the resonant frequency of the mode.[20] For ADMX, the empty-cavity frequency ($f_0$) for the TM$_{010}$ mode is ~550 MHz, and the theoretical unloaded quality factor $Q_0 \approx 500{,}000$ at 4 K.

The frequency of the mode excited by the axion-photon conversion is $f \approx m_a c^2/h$, where $h$ is Planck's constant. Since the axion mass is unknown, ADMX must tune the cavity resonant frequency to carry out its search. ADMX adjusts the frequency of the modes using two conducting rods that run parallel to the longitudinal axis of the cavity.[20] The rods are independently repositioned to various radial distances from the centerline of the resonator, altering the boundary conditions and changing the electromagnetic field solutions. This tuning can be shown by solving the two-dimensional wave equation of a resonant cavity

$$\left(\nabla_t^2 + \gamma^2\right)\psi = 0. \qquad (6)$$

For TM$_{0n0}$ modes, $\psi = E_z$ and

$$\gamma^2 = 4\pi^2 \mu_0 \varepsilon f^2, \qquad (7)$$

where $\mu_0$ and $\varepsilon$ are the permeability and permittivity within the cavity, respectively. Because the Laplacian is a measure of scalar curvature and the additional boundary conditions

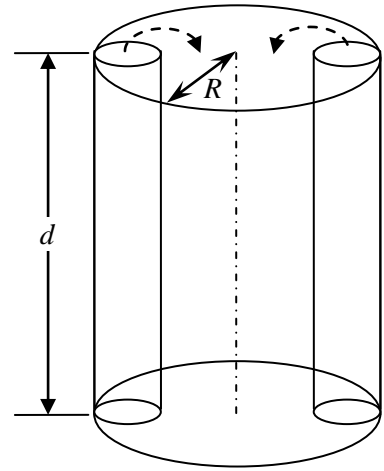

FIG. 2: Schematic of ADMX microwave cavity. The tuning rods are shown in the outmost configuration and move along the circular arcs. $R$ is the radius of the cavity and $d$ is the height.

increase the curvature of $E_z$, adding the conducting rods increases the frequency.

The frequency of the TM$_{010}$ mode increases as the rods approach the center of the cavity. The mechanical system moves the rods in a circular arc from near the wall to near the center. The experiment continually tunes the resonator in increments of $\sim f/5Q$. Figure 2 shows a schematic of the cavity.

ADMX has conducted searches for a number of years, successfully excluding axions with KSVZ coupling strength in the mass range of 1.9–3.5 μeV (460–860 MHz).[21-24] Figure 3 shows the published limits for the program. ADMX began searching the TM$_{020}$ mode range in the year 2014, extending their search up to approximately 6.2 μeV (1.5 GHz). However, the cavity design using two tuning rods does not provide for successful scanning above ~1.5 GHz. Increasing the diameter of the tuning rods can drive the TM$_{0n0}$ frequencies above 1.5 GHz, but the form factor ($C$) of the modes and scanning range of the cavity decrease significantly. New resonant cavity technologies need to be developed to enable searching for axions of greater mass.

The ADMX High-Frequency (ADMX-HF) is a similar but separate detector program that began operation in 2015. It uses a cavity with a volume of ~2x10⁻³ m³, and a single tuning rod with a radius of $r = R/2$. The detector search for axions at frequencies approximately one order of magnitude higher than ADMX. The TM$_{010}$ empty-cavity frequency for ADMX-HF is ~2.3 GHz and the theoretical unloaded $Q_0 \approx 200{,}000$ at 4 K.

## III. CAVITY DESIGN STUDY

The following relationships illustrate the dilemma faced by the axion search. The frequency of the exited mode scales approximately as the axion mass. The TM$_{010}$ frequency scales inversely with the cavity radius; the volume as the cube of the radius and $Q_0$ as the square root of the radius (at room temperature) when the ratio of the height to the radius, $d/R$, is maintained (which is desired to minimize mode crowding).[20]

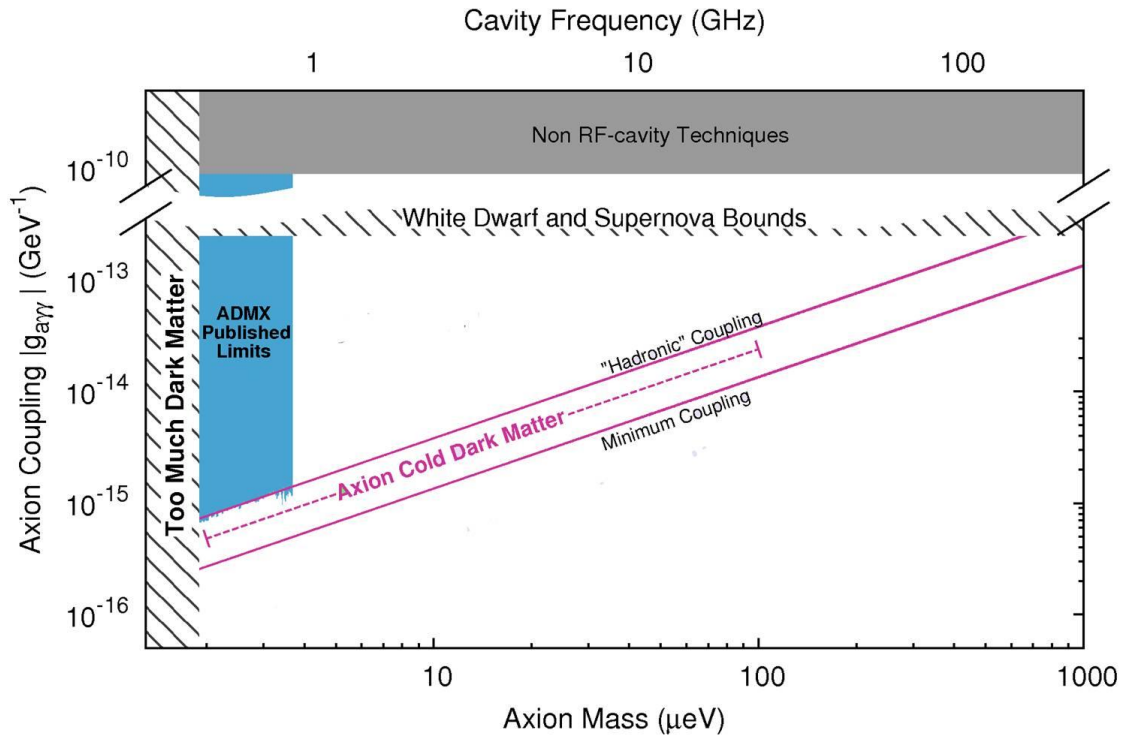

FIG. 3: Published limits for ADMX. The KSVZ limit is titled "'Hadronic' Coupling" and the DFSZ limit is titled "Minimum Coupling". The region in blue is excluded at 90% confidence assuming $\rho_a = 0.45$ GeV/cm. (figure courtesy of G. Rybka, ADMX, 2012)

From Eqs. (2), (4), and (5), we find the signal power scales as $\sim m_a V Q_0 \approx f^{-5/2}$.

At very low temperatures, the mean free path of the electrons in the cavity walls exceeds the cavity skin depth. The long mean free path puts the cavity in the anomalous skin effects (ASEs) regime,[20] where the skin depth scales as $f^{-1/3}$, $Q_0$ scales as $f^{-2/3}$, and the power will scale as $\sim f^{-8/3}$. Note, the normalization method of this study discussed later ensures that the results of the study are valid at all temperatures (see below).

Equation (2) shows the most efficient way of achieving cosmological sensitivity is to utilize a cavity with the maximum volume. However, as the frequency of the $TM_{0n0}$ mode scales as $1/R$ for an empty cylindrical cavity, searching for axions at the higher mass range would require bundling many smaller cavities together or using multiple metallic inserts in a single cavity. This paper represents a parametric study for the latter strategy.

The study compares the sensitivity of various cavity designs to that of a single empty cavity at the same frequency (reference cavity) to quantify its merit for CDM axion searches. The final results shown reflect only those designs that demonstrated a gain in sensitivity over the reference cavity throughout the frequency scan range. This evaluation technique provides a value for which other frequency-increasing methods, to include bundling multiple cavities, can be compared.

Two modified tuning techniques were investigated: periodic post arrays and regulating vanes. "Periodic post array" is used to describe any cavity that uses periodically spaced metallic posts. Periodic post arrays of 4, 5, 6, 7, and 8 posts, with assorted configurations, were evaluated. The "regulating vanes" concept uses movable partitions to define cells in the cylinder cross section that resonate in-phase in a TM mode. Regulating vanes with 4, and 6 vanes were studied.

Figure 4 shows top view schematics of examples of a post and a vane configuration. Note that in all designs, the many components move in concert. This requirement simplifies the drive mechanism that affects the motion and maintains angular symmetry.

Multiple rotating and translating post patterns were analyzed for a total of 99 tuning method designs (model cavities). Due to implementation constraints, translational configurations were not investigated in depth. Further, many configurations yielded a sensitivity that was lower than that of an empty cavity at the same frequency (reference cavity) during at least part of their frequency scan range. Table I shows only the geometry of the model cavity configurations that produced the best results. All designs in Table I use the rotating method to tune the model cavity, with the moving posts or vanes rotating about the central axis of the cavity as illustrated in Fig. 4.

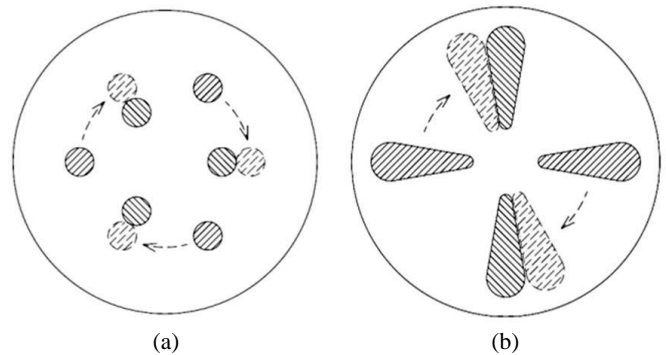

FIG. 4: Schematic drawings of sample tuning methods. (a) 6 post (3 moving) periodic post array resonator. (b) 4 regulating vane (2 moving) design. The arrows indicate the motion of the rotating post/vanes from solid hatching (maximum $TM_{010}$ frequency) to the dash hatching (minimum $TM_{010}$ frequency).

TABLE I: Scalable geometry of best performing model cavity designs. R_MOV/R is the radius of the location of the moving posts divided by the radius of the cavity; R_STAT/R is the radius of the location of the stationary posts divided by the radius of the cavity. POST DIA/R is the posts' diameter divided by the cavity radius. R_INNER/R is the radius of the location of the inner vane rounds divided by the radius of the cavity; R_OUTER/R is the radius of the location of the outer vane rounds divided by the radius of the cavity. Vane angle is the angle between the vane sides (in $\pi$ radians).

| Periodic post array | | | | | |
|---|---|---|---|---|---|
| Design | No. of stationary | No. of moving | R_MOV/R | R_STAT/R | POST_DIA/R |
| A | 2 | 2 | 0.413 | 0.413 | 0.350 |
| B | 2 | 2 | 0.425 | 0.425 | 0.300 |
| C | 2 | 2 | 0.454 | 0.454 | 0.350 |
| D | 3 | 3 | 0.375 | 0.625 | 0.250 |
| E | 3 | 3 | 0.375 | 0.750 | 0.375 |
| F | 3 | 3 | 0.400 | 0.675 | 0.275 |
| G | 3 | 3 | 0.400 | 0.800 | 0.300 |
| H | 4 | 3 | 0.450 | 0.450 | 0.200 |
| J | 4 | 3 | 0.525 | 0.525 | 0.175 |
| K | 4 | 4 | 0.550 | 0.550 | 0.200 |
| Regulating vane | | | | | |
| Design | No. of stationary | No. of moving | R_INNER/R | R_OUTER/R | Vane angle ($\pi$) |
| L | 2 | 2 | 0.250 | 0.800 | 0.150 |
| M | 2 | 2 | 0.300 | 0.700 | 0.150 |
| N | 3 | 3 | 0.300 | 0.600 | 0.050 |

Table I lists post location and size as a fraction of the cylinder radius. The TM frequencies scale as $1/R$. To the accuracy of the calculations, the form factor $C$ is independent of $R$. For $Q$, we take $R = 5$ cm; one must adjust $Q$ as $R^{1/2}$ for normal skin effect or $R^{2/3}$ for ASE. The figure of merit calculations is valid for any temperature and radius.

The numerical analysis was conducted with a commercially available three-dimensional finite element modeling program (COMSOL version 4.2). All model cavities had a radius of 5 cm and a height of 25 cm. The posts and vanes were assumed to run the entire length of the resonator without any air gaps. A typical model consisted of ~20K tetrahedral mesh elements and used impedance boundary conditions to determine the power loss in the copper walls and tuning rod or vanes.

Because the $TM_{010}$ mode has the largest coupling form factor ($C_{mnp}$), the study used that mode as the benchmark for evaluating various designs. To make the analysis applicable for cavities of any radius, the results are normalized such that they are not a function of the cavity radius, yielding scalable results. The frequencies ($f$) obtained from the simulation were multiplied by the radius of the model cavity ($R = 5$ cm) to produce the scalable parameter, $\Lambda$. The frequency of a fabricated cavity with an equivalent tuning configuration would be the following:

$$f_{cavity} = \frac{\Lambda}{R_{cavity}}, \tag{8}$$

where the subscript "cavity" denotes the values of the fabricated (not model) cavity. For example, a model cavity design with a (scalable) frequency range of $\Lambda = 0.2\text{-}0.3$ GHz-m would yield a frequency range of $f = 0.8\text{-}1.2$ GHz for a fabricated cavity of radius 0.25 m. The geometry of the posts/vanes would also need to be scaled by the radius of the fabricated cavity, and $Q$ would need to be adjusted.

Ideally, this study would focus on minimizing the total scan time of the experiment to reach a given axion coupling sensitivity over a broad mass range. However, total search time is not a feasible parameter to use as an optimizing variable for the study as it requires the piecewise integration of the entire scan range of each design, which would be extremely time consuming to complete for numerous designs. Additionally, dissimilar scan ranges would inherently have differing scan times with normalization being difficult to define, making comparing various designs problematic. Accounting for dissimilar scan ranges of designs would require a more complex systematic study that would involve multiple cavity designs and take into account the time required to exchange cavities, which is beyond the scope of this study.

To evaluate the merit of each design, three frequency dependent parameters were computed throughout the scan range: quality factor, form factor ($C$), and scan rate. The quality factor is defined as

$$Q \equiv -\frac{2\pi f}{\frac{dU}{dt}} U, \tag{9}$$

where $U$ is the energy stored in the cavity.[25] The scan rate, among other parameters, determines the maximum speed at which the search can be conducted for a given signal-to-noise ratio ($s/n$).

The Dicke radiometer equation relates the signal-to-noise ratio to power,

$$\frac{s}{n} = \frac{P}{k_B T_n} \sqrt{\frac{t}{b}} \qquad (10)$$

where $T_n$ is the system noise temperature, $k_B$ is Boltzmann's constant, $t$ is the integration time, and $b$ is the signal bandwidth.[26] For axion searches, typically $s/n > 5$. The signal bandwidth for a CDM axion detector comes from the velocity dispersion of the axion,

$$b = \frac{m_a}{Q_a}, \qquad (11)$$

where $Q_a \approx c^2/v\Delta v$ is the "quality factor" of axions, such that $v$ and $\Delta v$ are the velocity and velocity dispersion, respectively, of the axions on earth. Combining (10) and (11) and solving for $t$ yields

$$t = \left(\frac{s}{n}\right)^2 \frac{k_B^2 T_n^2}{P^2} \frac{m_a}{Q_a}. \qquad (12)$$

If the quality factor of the cavity ($Q$) is assumed to be less than the quality factor of the axion ($Q_a$), then the number of signal bandwidths that can be scanned simultaneously by a cavity is

$$N = \frac{Q_a}{Q_L}. \qquad (13)$$

where $Q_L$ is the loaded quality factor of the cavity (assumed critically coupled). The time ($\Delta t$) required to scan over a small frequency range ($\Delta f$) is

$$\Delta t = \frac{\Delta f}{Nb}\left(\frac{s}{n}\right)^2 \frac{k_B^2 T_n^2}{P^2} \frac{m_a}{Q_a}. \qquad (14)$$

Combining Eq. (2) with Eq. (14) and rearranging yields the scan rate[20,27]

$$\frac{df}{dt} \approx \left(\frac{s}{n}\right)^{-2} \left(\frac{1}{k_B T_n}\right)^2 \frac{g_{a\gamma\gamma}^4 \rho_a^2}{m_a^2} B_0^4 V^2 C^2 Q_L Q_a. \qquad (15)$$

The scan rate provides a measure of the sensitivity of the experiment, defined by the weakest axion-photon coupling for which the experiment can detect a signal in a given time. From Eq. (15), the sensitivity of a detector at a specified confidence level is found to be

$$g_{a\gamma\gamma}(f) \propto \left(\frac{df}{dt}\right)^{\frac{1}{4}}. \qquad (16)$$

Equation (16) indicates that doubling the sensitivity of the detector is equivalent to reducing the minimum scanning time at a given frequency by a factor of 16, and illustrates the difficulty of detecting CDM axions at the DFSZ limit. The coupling sensitivity as a function of frequency was selected for the basis of comparing model cavity designs because the parameter directly correlates to the published limits for axion searches (see y-axis of Fig. 3).

When evaluating different model cavities within the same magnetic field at a given frequency, only the form factor and quality factor vary in Eq. (15). To establish a non-dimensional figure of merit, the results of each model cavity simulation were normalized with equivalent results of an empty cylindrical cavity with the identical $TM_{010}$ frequency. The $TM_{010}$ frequency of the reference cavity is a function of its radius only, given by

$$f_0 = \frac{x_{01}c}{2\pi R_0} = 0.115\,\text{GHz}\left(\frac{1\,\text{m}}{R_0}\right), \qquad (17)$$

where $R_0$ is the radius of the reference cavity and $x_{01}$ is the first root of the zeroth order Bessel function ($J_0$).[25] The height of the reference cavities were kept the same as the model cavities ($d = 25$ cm). Note, the aspect ratio (ratio of the height to the radius) of the reference cavities ($d/R_0$) changes as a function of frequency, while the aspect ratio for the model cavities in this study are held constant ($d/R = 5$).

The resulting figure of merit used to evaluate all model cavity configurations was

$$\mathcal{F}(f) \equiv \left[\frac{V^2(C_{010})^2 Q}{V_0^2 C_0^2 Q_0}\right]^{\frac{1}{4}}, \qquad (18)$$

where the numerator uses the volume, form factor, and quality factor of the model cavity, and the denominator uses the volume, form factor, and quality factor of the reference cavity (denoted by the subscript 0). For the $TM_{010}$ mode of the reference cavity, $C_0 = 0.692$. The figure of merit for a cavity is a function of frequency, because the form factor and quality factor of the model cavity, and the volume and quality factor of the reference cavity vary with frequency.

The figure of merit is dominated by the volume ratio $V/V_0 = (A/A_0)^2$, where $A_0 = 0.115$ GHz-m for the $TM_{010}$ mode. This effect captures the intent of the study to produce large-volume cavities with higher TM mode frequencies. Note that $V$ and $C_0$ are constants, and $V_0$ and $Q_0$ are computed directly from the frequency. Only $C$ and $Q$ of the model cavity are obtained from the FEM simulation.

The normalization of $\mathcal{F}$ maintains scalability of the results as long as the aspect ratio of a fabricated cavity is the same as the model cavity. From Eq. (17), the radius of the reference cavity ($R_o$) can be written as a function of the model cavity radius ($R$)

$$R_0 = \frac{R}{A}\frac{x_{01}c}{2\pi}. \qquad (19)$$

Using Eq. (19), the ratio of volumes can be rewritten as the scalable function

$$\frac{V}{V_0} = \left(\frac{2\pi A}{x_{01}c}\right)^2. \qquad (20)$$

The form factor, $C_{mnp}$, is a scalable parameter by definition.

The quality factor for the $TM_{010}$ mode of a circular cylinder cavity is[25]

$$Q = \frac{d}{\delta}\frac{1}{1 + \xi_{01}\frac{Sd}{2A}}, \qquad (21)$$

where $\delta$ is the skin depth, and $S$ and $A$ are the cross-sectional circumference ($2\pi R$) and area ($\pi R^2$) of the cavity, respectively. The variable $\xi_{01}$ is a dimensionless number of order unity,

defined by

$$\zeta_{mn} \equiv \frac{A R_0^2}{x_{mn}^2 S} \frac{\sum_s \oint dl \left| \frac{\delta \psi_{mn}}{\delta n} \right|^2}{\int da \left| \psi_{mn} \right|^2}, \tag{22}$$

where $\psi_{mn}$ is the TM$_{nm0}$ solution to the field equations. $\delta \psi_{mn}/\delta n$ is the normal derivative of the field at the cavity surface. The closed line integral in the numerator is around each cavity surface, and the summation is over all (internal) cavity surfaces except the ends. Note, the ends are accounted for by the 1 in the denominator of (21). $\zeta_{01} = 1$ for an empty cavity.

Because the model and reference cavities are operating at the same frequency and temperature, their skin depth is identical. Noting $S/2A = 1/R$ for a circular cross-section, the ratio of quality factors can be expressed as

$$\frac{Q}{Q_0} = \frac{1 + \frac{\alpha R}{R_0}}{1 + \zeta_{01}\alpha}, \tag{23}$$

where $\alpha = d/R$ is the aspect ratio of the model cavity (see Fig. 2). $\zeta_{01}$ in the denominator is that of the model cavity, which can be rewritten using (19) as

$$\zeta_{01} = \frac{c^2 R^3}{8\pi^2 A^2} \frac{\sum_s \oint dl \left| \frac{\delta \psi_{01}}{\delta n} \right|^2}{\int da \left| \psi_{01} \right|^2}. \tag{24}$$

The field solution is always scalable, so the integrals in the numerator are proportional to $1/R$ and the integral in the denominator is proportional to $R^2$. $\zeta_{01}$ is therefore observed to be a scalable value, with the numerator and denominator both proportional to $R^2$. Utilizing Eq. (19), we can rewrite Eq. (23) as a scalable equation (except for the aspect ratio),

$$\frac{Q}{Q_0} = \frac{1 + \alpha \frac{\Lambda}{K_{01}}}{1 + \zeta_{01}\alpha}, \tag{25}$$

With

$$K_{01} \equiv \frac{x_{01} c}{2\pi}. \tag{26}$$

Equations (19)–(26) show that the figure of merit ($\mathcal{F}$) is a scalable value except for the cavity aspect ratio ($\alpha$). The results of this study are valid for any cavity with an aspect ratio of $d/R \approx 5$. From Eq. (25), we derive the result for the figure of merit due to varying the aspect ratio as

$$\mathcal{F}_2 = \mathcal{F}_1 \left[ \left( \frac{\frac{K_{01}}{\Lambda} + \alpha_2}{1 + \alpha_2\zeta_{01}} \right) \left( \frac{1 + \alpha_1\zeta_{01}}{\frac{K_{01}}{\Lambda} + \alpha_1} \right) \right]^{\frac{1}{4}}, \tag{27}$$

where $\alpha_i$ is the aspect ratio of the respective cavities. To convert the figures of merit from this study to that of a cavity with a differing aspect ratio, $\alpha_1 = 5$, $\alpha_2$ is the aspect ratio of the fabricated cavity, $\mathcal{F}_1$ is the figure of merit at a give frequency from the study, and $3 \lesssim \zeta_{01} \lesssim 7$.

The normalization also accounts for modifications of $Q$ due to temperature. At low temperatures, the conductivity of copper improves which increases $Q$, and the copper surfaces enter the anomalous skin depth regime which eventually stops $Q$ from further increasing.[28] The $Q$ of both the model cavity and the reference cavity ($Q_0$) would change proportionately, keeping their ratio a constant. The antenna coupling is also accounted for in the normalization as the normalization cavity always has the same coupling as the model cavity, resulting in $Q_L/Q = Q_{L9}/Q_0$. Thus, the results of the study are valid at any detector temperature and for any cavity size with $\alpha \approx 5$.

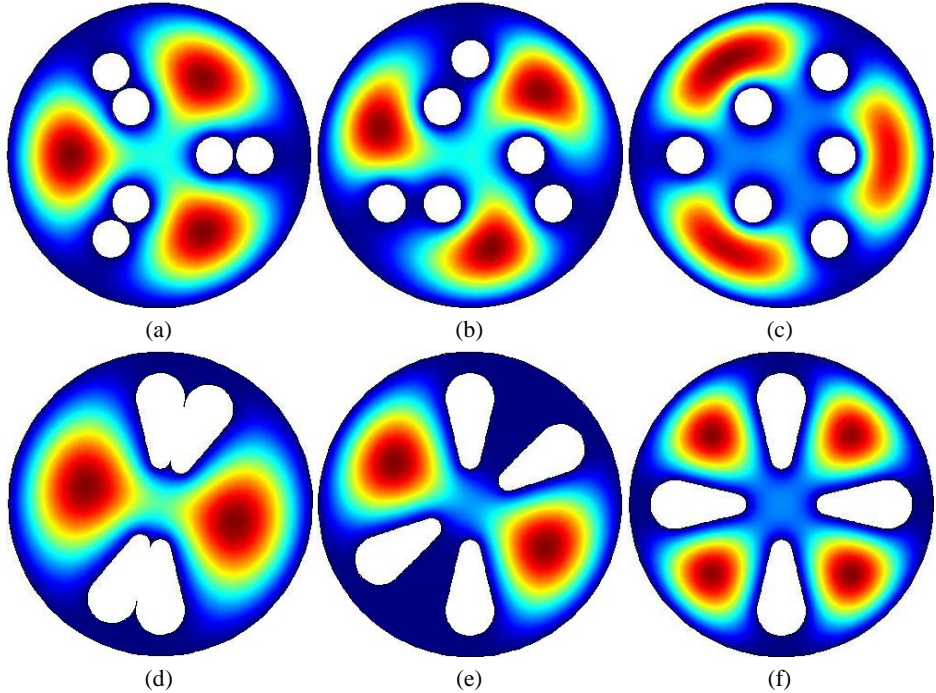

FIG. 5: Electric field magnitude for the TM$_{010}$ mode of sample cavity designs. (a)–(c) 6-post periodic post array resonator from the minimum (left) to maximum (right) frequency configurations. (d)–(f) 4 regulating vane design from the minimum (left) to maximum (right) frequency configurations. The white areas represent the frequency-control posts or vanes (tuning devices).

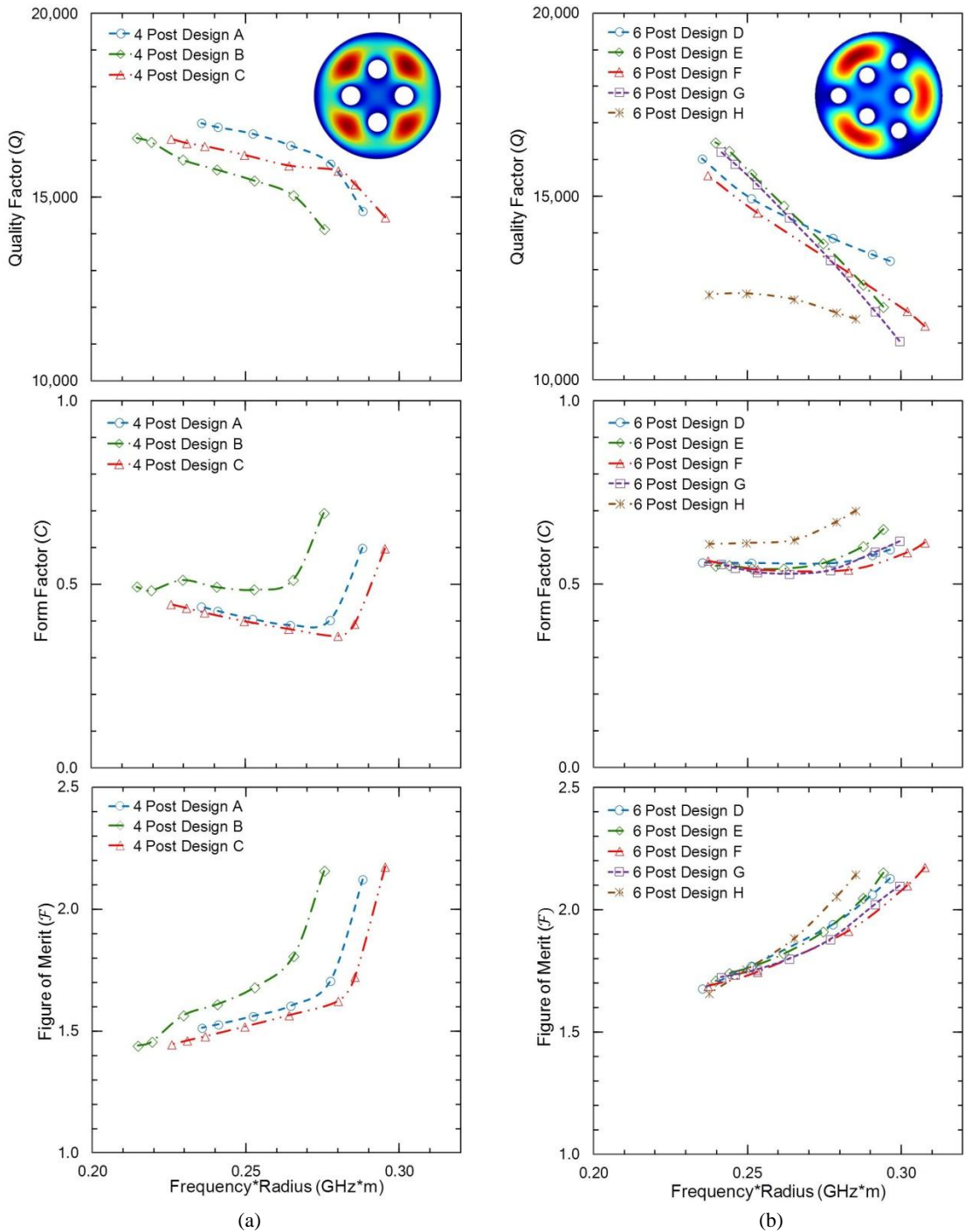

FIG. 6: Unloaded quality factor (top), form factor (middle), and figure of merit (bottom) for select 4-post (a) and 6-post (b) cavity designs (see Table I). $Q$ values shown are for an aspect ratio of $\alpha = 5$, and a temperature of 300 K. The normalized value $\mathcal{F}$ is shown on the x-axis. The frequency range of the plot is 0.95-1.5 GHZ for ADMX ($R = 21$ cm) and 4.0-6.4 GHz for ADMX-HF ($R = 5$ cm). The insets show the electric field of the $TM_{010}$ mode for the 4-post and 6-post designs at the highest frequency configuration for visualization.

For visualization, Fig. 5 shows a cross section (normal to the cylinder axis) of the electric field of the $TM_{010}$ mode for two model cavities similar to those depicted in Fig. 4. Typically, a more evenly dispersed field will result in a higher form factor and thus a higher figure of merit.

The study demonstrated symmetry breaking of a cavity configuration, resulting in transverse mode-localization, where the bulk of $E_z$ in the $TM_{010}$ mode is found in the cell(s) with the largest cross-sectional area. This is most obvious in Fig. 5(e), where the field is divided into two segments, and there are two distinct areas of the cavity that have approximately no field. Rotating the tuning devices towards the localized fields further "squeezes" the fields, which causes an increase in frequency. The squeezing continues until a new symmetry arises from the configuration, and the mode significantly changes, as seen in Fig. 5(f).

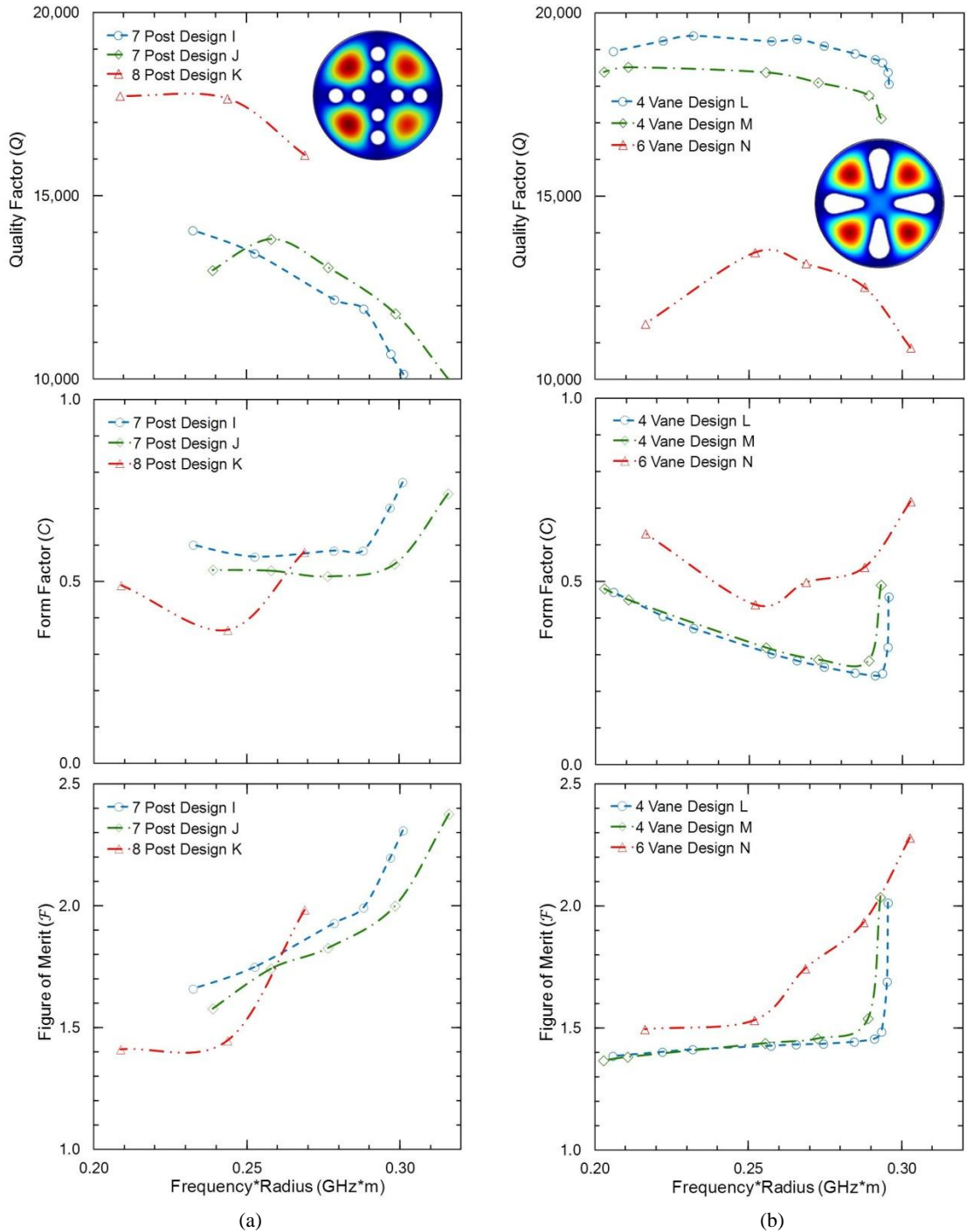

FIG. 7: Unloaded quality factor (top), form factor (middle), and figure of merit (bottom) for select 7-post and 8-post (a), and 4-vane and 6-vane (b) cavity designs (see Table I). $Q$ values shown are for an aspect ratio of $\alpha = 5$, and a temperature of 300K. The normalized value $\Lambda$ is shown on the x-axis. The frequency range of the plot is 0.95-1.5 GHz for ADMX ($R = 21$ cm) and 4.0-6.4 GHz for ADMX-HF ($R = 5$ cm). The insets show the electric field of the $TM_{010}$ mode for the 8-post and 4-vane designs at the highest frequency configuration for visualization.

Mode localization in periodic post arrays has been observed previously,[20,29] and lateral localization has been shown to be predictable from the eigenfunction boundary problem of Eq. (6).[30,31] However, the results from this study show that the mode configuration is notably sensitive to symmetry breaking and local cross-sectional areas.

The resulting quality factor, form factor, and figure of merit for the selected "higher performing" cavities are shown in Figs. 6 and 7 as a function of $\Lambda$. A cavity height of $d = 25$ cm was used to compute the unloaded $Q$ for comparison. The results shown are for an aspect ratio of 5. The aspect ratios of the ADMX and ADMX-HF resonators are 4.8 and 5.0, respectively. Refer to Table I for scalable geometry of the model cavity designs.

The quality factor for many of the model cavity designs decreased as the frequency increases. This may seem counterintuitive, since the skin depth decreases as $1/\sqrt{f}$ in the normal skin-effect regime. This phenomenon can be explained using Eq. (24). The tuning devices squeeze $E_z$ as the mode frequency increases, as discussed above. The squeezing decreases $\int da \, |\psi_{01}|^2$ while increasing $\delta\psi_{01}/\delta n$, which increases $\xi_{01}$. This effect dominates the gain from the reduction in skin depth, resulting in lowering of $Q$ as the frequency increases. The squeezing effect can also be interpreted a change to the effective volume to surface area ratio. Note that $Q$ for an empty cavity is proportional to $f^{-2/3}$ in ASE.

The loss in $Q$ with frequency is most noticeable in the 6-post designs, where the squeezing of the electric field is most visible (see Figs. 5(a)–5(c)). $Q$ is almost constant with frequency for the 4-vane designs, and the squeezing of the field is less obvious. The rapid drop in the $Q$ of the 4-vane cavity near the highest frequency is due to the field configuration changing from two cells (see Fig. 5(e)) to four cells (see Fig. 5(f)), which causes a rapid increase in the summation of $\delta\psi_{01}/\delta n$ with a small increase in frequency.

In general, the form factor of a cavity grows as $E_z$ fills a larger area of the cross section. This often occurred at the frequency extremes of the cavities because, at the extremes, the field would maximize its area in differing configurations (see Fig. 5). During the frequency transitions, the field occupies a smaller area, causing $C$ to be smaller at the mid-frequency configurations.

The phenomenon is most noticeable in the 4-vane designs. In the minimum frequency configuration (Fig. 5(d)), $E_z$ is separated into 2 cells and occupies most of the cavity volume around the vanes. During the frequency transition (Fig. 5(e)), the two cells are squeezed and there are noticeable locations where no field is present. The unfilled sections cause the form factor to decrease. Near the maximum frequency configuration (Fig. 5(f)), the field separates into 4 cells, filling the volume of the cavity rapidly with the configuration change. This causes the sharp increase in $C$ at the high-frequency end of the 4-vane cavities.

The 4-post, 7-post, 8-post, and 6-vane designs also demonstrate the phenomenon to varying amounts, but the 6-post design does not. This is because the $E_z$ field of the 6-post cavity maintains 3 sections throughout the frequency range, avoiding any sudden change in field area. The study showed that the highest $\mathcal{F}$ most commonly resulted in cavity designs that maximized $C$ at the frequency extremes. Note that many of the designs that were not selected as the "best" did not have high fill volumes at the extremes, resulting in a less-than-optimum design.

Small local increases in the $Q$ and $C$ curves are typically a result of a field transition. However, transition points can be challenging for field simulations and convergence difficulties may exaggerate the bumps to some extent.

The figure of merit is derived directly from the computed $Q$ and $C$ values using Eq. (18). Because $\mathcal{F}$ is proportional to $C^{1/2}$ and $Q^{1/4}$, the form factor dominates $\mathcal{F}$. The figure of merit also increases with frequency because the volume of the reference cavity, $V_0$, is proportional to $1/f^2$. This effect mirrors the scan rate proportionality of $g_{a\gamma\gamma}^4/m_a^2$ in Eq. (15).

Combining the power from multiple, equivalent cavities in-phase increases the signal power by the number of cavities. The number of identical empty cavities needed to equal the searching capability of the model cavities in the study would be $\mathcal{F}^2$ when combined coherently (frequency locked), and $\mathcal{F}^4$ when combined incoherently. So, with a figure of merit of 2, a multi-post/vane cavity has the equivalent searching ability of 4 empty cavities with equivalent height combined in phase, not accounting for any broadening of the bandwidth due to frequency mismatch or power loss due to phase incoherence. Conducting a search with multiple empty cavities is not practical, as they cannot be tuned, but this comparison is useful to show equivalency.

The results show that both tuning methods significantly increased the TM$_{010}$ frequency while maintaining reasonable $\mathcal{F}$ values. In general, the periodic post array designs performed moderately better than the regulating vanes. While the number of posts has a noticeable impact on performance, there was marginal gain recovered by optimizing the geometry of a specific design. No single point design showed a distinct improvement in sensitivity. Cavities with the most evenly distributed electric field throughout the frequency tuning range maintained the highest $\mathcal{F}$.

## IV. CONCLUSION

This cavity design study showed that a significant increase in tuning range with acceptable detection parameters was achievable with periodic post array resonators and regulating vanes. While there were notable differences among the configurations simulated, modest improvements in sensitivity could be obtained from many designs. Both design configurations (posts or vanes) could be used to develop microwave cavities for dark matter axion searches with a haloscope detector.

Several designs demonstrated a sensitivity gain of approximately 1.5-2.0 over that of a single empty cavity, which translates into a factor of 5-16 improvement in the scan rate. To match the searching capabilities of the higher-performing cavities depicted in the results, an axion scan would need to combine the power from approximately four cavities of equivalent length with one small tuning rod when frequency locked or approximately eight such cavities incoherently.

No single design proved far superior to all others. Choices then fall back on cost of fabrication, and sensitivity of frequency range and $\mathcal{F}$ to misalignments. Typically, posts are less expensive to fabricate than vanes due to their simplistic design. Vanes also have an orientation requirement that posts do not, increasing the sensitivity and complexity of assembly of the regulating vanes over that of the periodic post array resonators.

## ACKNOWLEDGEMENTS

I.S. acknowledges support by the U.S. Department of Defense through the National Defense Science and Engineering Graduate Fellowship Program and the National Aeronautics and Space Administration through the Florida Space Research Program. This work was supported in part by the Department of Energy under Grant No. DE-SC0010280 at the University of Florida and under Contract No. DEAC52-07NA27344 at Lawrence Livermore National Laboratory, and in part by the National Science Foundation under Grant No. PHY-1306729 at the University of California Berkeley.